\documentclass[aps,preprint,floatfix,nofootinbib,showpacs]{revtex4}
\usepackage{graphicx}
\newcommand{\U}{{\cal U}}
\begin{document}

\title{Unparticle effects in photon-photon scattering}
\renewcommand{\thefootnote}{\fnsymbol{footnote}}
\author{ 
Chun-Fu Chang$^{1}$,
Kingman Cheung$^{1,2}$, 
and Tzu-Chiang Yuan$^{2}$
 }
\affiliation{$^1$Department of Physics, National Tsing Hua University, 
Hsinchu 300, Taiwan
\\
$^2$Physics Division, National Center for Theoretical Sciences,
Hsinchu 300, Taiwan
}

\renewcommand{\thefootnote}{\arabic{footnote}}
\date{\today}

\begin{abstract}
Elastic photon-photon scattering can only occur via loop diagrams in
the standard model and is naturally suppressed. Unparticle can induce
tree-level photon-photon scattering through the operator $F_{\mu\nu}
F^{\mu\nu} O_\U$ for spin-0 unparticle or $F_{\mu\alpha}
F^{\alpha}_{\nu} O^{\mu\nu}_\U$ for spin-2 unparticle. 
Due to the peculiar CP-conserving phase $\exp(-i d_\U \pi)$
associated with the $s$-channel unparticle propagator, its
interference effects with the $t$- and $u$-channels on the total cross
section and the angular distribution are found to be some
significance.  In addition, we show that the cross sections via
unparticle exchange can be substantially larger than the standard
model contribution.
\end{abstract}


\maketitle

\section{Introduction}

Recently, Georgi \cite{unparticle} pointed out an interesting 
possibility for the existence of a scale-invariant sector with a 
continuous mass distribution. 
This scale invariant stuff was coined the term
``unparticle'' to describe a possible scale-invariant hidden sector
sitting at an infrared fixed point at a high scale $\Lambda_\U$.
The scale-invariant sector may be 
weakly or strongly interacting but its effects
on the standard model (SM) is assumed to be weakly interacting.
In Georgi's scheme \cite{unparticle},
the hidden sector communicates with the SM content via a messenger
sector characterized by a high mass scale $M$.  At energy below $M$,
one can integrate out the messenger sector and ends up with
effective operators suppressed by inverse powers of $M$ in the
following form
\begin{eqnarray}
\label{effectiveop}
\frac{1}{M^{d_{SM} + d_{UV} - 4}} {\cal O}_{SM} {\cal O}_{UV} \;,
\end{eqnarray}
where ${\cal O}_{SM}$ and ${\cal O}_{UV}$ represent local operators of
the SM and hidden sector with scaling dimensions $d_{SM}$ and $d_{UV}$,
respectively.  As one scales down the theory from $M$, the hidden
sector may flow to an infrared fixed point at the scale $\Lambda_\U$,
which is generated by quantum effects via dimensional transmutation
for example.  At the fixed point the hidden sector becomes scale
invariant, the above operator Eq. (\ref{effectiveop}) has to be
replaced by a new set of operators of similar form
\begin{eqnarray}
\label{unparticleop}
C_{{\cal O}_{\U}} \frac{\Lambda^{d_{UV} - d_{\U}}_\U}{M^{d_{SM} + d_{UV} - 4}}
 {\cal O}_{SM} {\cal O}_{\U}\;,
\end{eqnarray}
where ${\cal O}_{\U}$ is the unparticle operator with a scaling
dimension $d_\U$ and $C_{{\cal O}_{\U}}$ is the unknown
coefficient. Due to the underlying theory is a scale invariant
interacting theory, the scaling dimension $d_\U$ needs not having the
canonical values of integer or half-integer like the free boson or
free fermion cases. Besides its scaling dimension, 
the unparticle operator ${\cal O}_{\U}$ can be
characterized by scalar, vector, tensor, or spinor etc according to its
Lorentz group representation.

Despite the scale invariant sector remains unspecified, the 2-point
function \cite{unparticle} and the Feynman propagator
\cite{unparticle-propagator,CKY} of the unparticle field operator
${\cal O}_\U$ can be determined by scale invariance.
The normalization of the 2-point function of unparticle operator
of scaling dimension $d_\U$ was fixed by Georgi \cite{unparticle} to be
the same as the phase space of $d_\U$ massless particles.  
On the other hand, the most peculiar feature of 
the unparticle propagator is a phase factor $\exp(-i d_\U \pi)$ 
associated only with time-like momenta.  
This CP-conserving phase has been shown to have interesting
interference effects at high energy experiments 
\cite{unparticle-propagator,CKY} and other phenomenology.

In this work, we consider photon-photon scattering via unparticle
exchanges.  The SM contribution to photon-photon scattering can only
arise from loop diagrams with all charged particle running around the loop 
and thus is highly suppressed. It is anticipated
that the cross section due to unparticle exchange can easily surpass
the SM cross section at high enough energies, because exchanges of unparticle
are at the tree-level.
Moreover, photon scatters via unparticle exchanges
in all $s$-, $t$-, and $u$-channels.  The peculiar phase $\exp(-i d_\U \pi)$
associated with the $s$-channel exchange gives rise to
interesting interference with the $t$- and $u$-channel amplitudes. 
Similar effects had been studied in the
model of large extra dimensions \cite{extra}.

Note that similar ideas for the spin-0 unparticle 
have been pursued recently in Refs. \cite{wis,nobu}.
However, our analytic results disagree with Ref. \cite{wis}.  We
suspect that the phase factor $\exp(-i d_\U \pi)$ associated with the
$s$-channel unparticle propagator was not taken care of properly.  Our
results are consistent with Ref. \cite{nobu} where we overlap.  In
addition, we extend these previous calculations to the
spin-2 unparticle exchange, which is highly nontrivial.

It has been pointed out recently by Grinstein {\it et al} \cite{Grinstein} that 
the vector and tensor unparticle propagators for a conformal invariant 
hidden sector differ from a scale invariant ones. 
Unitarity constraints \cite{Mack, Minwalla} on the scaling dimensions of the unparticle 
operators with conformal symmetry
are also emphasized in their work \cite{Grinstein}.
In this work, we follow the original Georgi's scheme by assuming just scale invariance 
in the derivation of the unparticle propagators. 
Integrating out the heavy messenger sector can also lead to contact interactions among 
SM fields of the form 
${\cal O}_{SM}{\cal O}^\prime_{SM}/M^{d_{SM} + d^\prime_{SM} - 4}$ and they can compete with the effects from unparticle exchanges \cite{Grinstein}. 
For example, the following two dimension 8 operators 
$(F^{\mu\nu} F_{\mu\nu})^2/M^4$ and 
$(F^{\mu\nu} F^{\alpha\beta} F_{\mu\alpha} F_{\nu\beta})/M^4$ 
can be induced and they can also contribute to the elastic photon-photon 
scattering. We assume the coefficients of these operators are minuscule and ignore 
them in our analysis. 
A complete analysis for the photon-photon scattering including all the interference effects among the SM contribution, unparticle exchanges as well as these contact interactions 
is interesting but beyond the scope of this work.

The organization of the paper is as follows.  In the next section, we
give in details the scattering amplitudes for $\gamma\gamma \to \gamma
\gamma$ via spin-0 as well as spin-2 unparticle exchange.  In
Sec. III, we compare the unparticle contribution with the SM
contribution in the angular distribution and in the total cross section.
We also look into the nontrivial effects of the phase $\exp(-i d_\U
\pi)$ of the $s$-channel propagator.  We conclude in Sec. IV.

\section{Photon-photon scattering}
 
The interaction of spin-0 unparticle $\U$ with the photon
can be parameterized by \cite{unparticle,CKY}
\begin{equation}
{\cal L}_{\mathrm{eff}}  \ni  \lambda_0 \, \frac{1}{\Lambda_\U^{d_\U}}\,
F_{\mu\nu} F^{\mu\nu}\, O_\U \;, 
\end{equation}
where $ \lambda_0 $ is an unknown coefficient of order $O(1)$, and
$F_{\mu\nu}$ is the field strength of the photon field.
The unparticle propagator is \cite{unparticle-propagator,CKY}
\begin{equation}
\label{unpropagator}
 \Delta_F (P^2) = \frac{A_{d_\U}}{2 \sin (d_\U \pi) } \, (-P^2)^{d_\U-2} \;,
\end{equation}
where $A_{d_\U}$ is given by
\[
    A_{d_\U}={16\pi^2\sqrt{\pi}\over (2\pi)^{2{d_\U}}}
       { \Gamma({d_\U}+{1\over
       2})\over\Gamma({d_\U}-1)\Gamma(2\,{d_\U})}  \;.
\]
The peculiar phase associated with the propagator arises from the
negative sign in front of $P^2$ in Eq.(\ref{unpropagator}):
\begin{equation}
\label{branchcut}
 (-P^2)^{d_\U -2}=\left \{
\begin{array}{lcl}
|P^2|^{d_\U -2}   & \quad & \hbox{if $P^2$ is negative and real, } \\
|P^2|^{d_\U -2} e^{-i d_\U \pi} & & \hbox{for positive $P^2$ with 
an infinitesimal $i0^+$} . 
\end{array} \right.
\end{equation}
Therefore, the $s$-channel propagator has the nontrivial phase
$\exp(-i d_\U \pi)$ while the $t$- and $u$-channel propagators do not.

There are three Feynman diagrams contributing to 
$\gamma(p_1)\;\gamma (p_2)\; \to \;\gamma(k_1)\; \gamma(k_2) $
with the unparticle exchanges in $s$-, $t$-, and $u$-channels. 
The sum of amplitudes for these three diagrams is given by
\begin{equation}
{\cal M} =  - 16 \lambda_0^2 Z_{d_\U} \, \frac{1}{\Lambda_\U^4} \,
 \left ( {\cal M}_s + {\cal M}_t + {\cal M}_u  \right )^{\mu\nu\rho\sigma} \,
\epsilon^*_\sigma (k_1)\,
\epsilon^*_\rho (k_2)\,
\epsilon_\nu (p_1)\,
\epsilon_\mu (p_2)  \;,
\end{equation}
where
\begin{eqnarray}
{\cal M}^{\mu\nu\rho\sigma}_s &=& 
\left( \frac{-s}{\Lambda_\U^2} \right )^{d_\U -2} \, 
\left( - k_1\cdot k_2 g^{\rho\sigma} + k_1^\rho k_2^\sigma \right )
\left( - p_1\cdot p_2 g^{\mu\nu} + p_1^\mu p_2^\nu \right ) \,,
\nonumber  \\
{\cal M}^{\mu\nu\rho\sigma}_t &=& 
\left( \frac{-t}{\Lambda_\U^2}  \right )^{d_\U -2} \, 
\left( k_2\cdot p_2 g^{\mu\rho} - k_2^\mu p_2^\rho \right )
\left( k_1\cdot p_1 g^{\nu\sigma} - k_1^\nu p_1^\sigma \right ) \,, 
 \nonumber \\
{\cal M}^{\mu\nu\rho\sigma}_u &=&  
\left( \frac{-u}{\Lambda_\U^2} \right )^{d_\U -2} \, 
\left( k_2\cdot p_1 g^{\nu\rho} - k_2^\nu p_1^\rho \right )
\left( k_1\cdot p_2 g^{\mu\sigma} - k_1^\mu p_2^\sigma \right ) \nonumber \;.
\end{eqnarray}
In the above amplitude, we can write the Mandelstam variables as 
\begin{equation}
(-s)^{d_\U-2} = s^{d_\U -2} e^{-i d_\U \pi}\;, \qquad
(-t)^{d_\U-2} = |t|^{d_\U -2} \;, \qquad
(-u)^{d_\U-2} = |u|^{d_\U -2} 
\end{equation}
such that the phase $\exp(-id_\U \pi)$ associates manifestly 
with the $s$-channel only.
It is obvious that each channel is separately gauge invariant.
The square of the amplitude averaged over initial polarizations is given by
\begin{equation}
\label{msqspin0}
\overline{\sum} |{\cal M}|^2 =  \frac{16 \lambda_0^4 Z^2_{d_\U}  }
 {\Lambda_\U^{4 d_\U}}\, \Biggl\{ 
  s^{2 d_\U} + |t|^{2 d_\U} + |u|^{2 d_\U} 
+ \cos (d_\U \pi) \left[ (s |t|)^{d_\U} + (s|u|)^{d_\U} \right]
+ (|t| |u|)^{d_\U} \Biggr\} \;.
\end{equation}
If the phase factor $\cos(d_\U \pi)$ were removed, the amplitude squared
would have been symmetric in $s \leftrightarrow t \leftrightarrow u$. 
Note that we have written the Mandelstam variables as $s,|t|, |u|$,
where $|t| = s(1 - \cos\theta)/2$ and $|u| = s( 1+\cos\theta)/2$ and
$\theta$ is the central scattering angle.
The angular distribution is given by
\begin{eqnarray}
\label{dsigmadcos}
\frac{d\sigma}{d \cos\theta} &=& \frac{1}{2}\frac{1}{ 32 \pi s} 
 \, \overline{\sum} |{\cal M}|^2 \nonumber \\ 
&=& \frac{ \lambda_0^4 Z^2_{d_\U} }{ 4 \pi \Lambda_\U^{4 d_\U} }\,
 s^{2 d_\U-1}\,
\Biggr\{
 1 + \left( \frac{1-\cos\theta}{2} \right)^{2 d_\U} 
   + \left( \frac{1+\cos\theta}{2} \right)^{2 d_\U}  \nonumber \\
&&
   + \cos(d_\U \pi) \left[ \left( \frac{1-\cos\theta}{2} \right)^{d_\U} 
                         + \left( \frac{1+\cos\theta}{2} \right)^{d_\U} 
                   \right ]
 + \left ( \frac{1 -\cos^2\theta}{4} \right )^{d_\U} 
\Biggr \} \;,
\end{eqnarray}
where the range of integration for $\cos\theta$ is from $-1$ to 1.
The total cross section can be obtained analytically in closed form 
by integrating Eq.(\ref{dsigmadcos}) over $\cos\theta$, viz.,
\begin{equation}
\sigma = \frac{ \lambda_0^4 Z^2_{d_\U} }{ 2 \pi \Lambda_\U^{4 d_\U} }
\, s^{2 d_\U-1}\, \biggr\{
 1 + \frac{2}{2 d_\U +1} + \frac{2 \cos(d_\U \pi)}{d_\U +1} 
   + \frac{\sqrt \pi}{2^{2 d_\U +1}} \frac{\Gamma(d_\U +1)}{\Gamma(d_\U + 3/2)}
\biggr \} \; .
\end{equation}

The effective interaction of 
spin-2 unparticle with the photon is given by \cite{unparticle,CKY}
\begin{eqnarray}
{\cal L}_{\mathrm{eff}}  \ni  \lambda_2 \, \frac{1}{\Lambda_\U^{d_\U}}\,
F_{\mu\alpha} F_\nu^{\alpha}\, O^{\mu\nu}_\U \;,
\end{eqnarray}
where $\lambda_2$ is an unknown effective coupling constant, of order
$O(1)$. 
Using the Feynman rules and the propagator derived in 
\cite{CKY}, the matrix element squared for 
elastic photon-photon scattering via spin-2 unparticle exchange
is found to be
\begin{eqnarray}
\label{msqspin2}
\overline{\sum} |{\cal M}|^2 & = & 
\frac{\lambda^4_2 Z^2_{d_\U}}{2 \Lambda_\U^{4 d_\U}}
\Biggl\{
s^{2 d_\U - 4} \left( t^4 + u^4 \right) 
+ \vert t \vert^{2 d_\U - 4} \left( s^4 + u^4 \right) 
+ \vert u \vert^{2 d_\U - 4} \left( s^4 + t^4 \right) \Biggr. \nonumber \\
&  & \; \Biggl. + 2 \cos \left( d_\U \pi \right) s^{d_\U - 2} 
\left[ 
\vert t \vert^{d_\U - 2} u^4 + \vert u \vert^{d_\U - 2} t^4
\right]
+ 2 \left( t u \right)^{d_\U - 2} s^4
\Biggr\} \; .
\end{eqnarray}
As $d_\U \to 2$, the above expression is proportional to 
$s^4 + t^4 + u^4$ which is the familiar result \cite{extra} for the spin-2 
Kaluza-Klein graviton exchange in the large extra dimensions model.
However, the $s$-channel unparticle exchange contains a CP-conserving
phase factor $\exp(-i d_\U \pi)$ that does not share with the $t$- and
$u$-channels.  Therefore, the expressions of Eq.(\ref{msqspin0}) and
Eq.(\ref{msqspin2}) for the matrix element squared contain the factor
$\cos( d_\U \pi)$ in the interference terms between $s$- and $t$- and
between $s$- and $u$-channels. This is a unique feature of the unparticle.

\section{Results}

\begin{figure}[t!]
\centering
\includegraphics[width=5in]{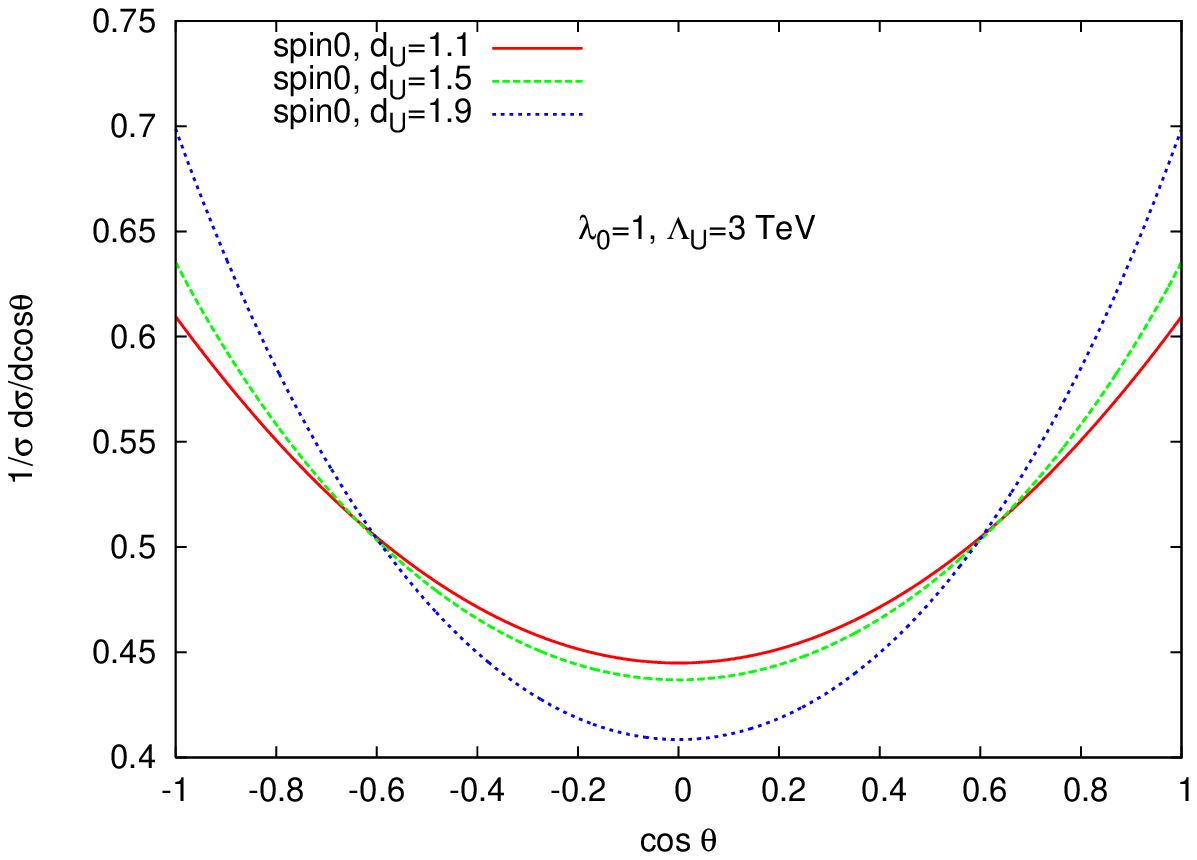}
\includegraphics[width=5in]{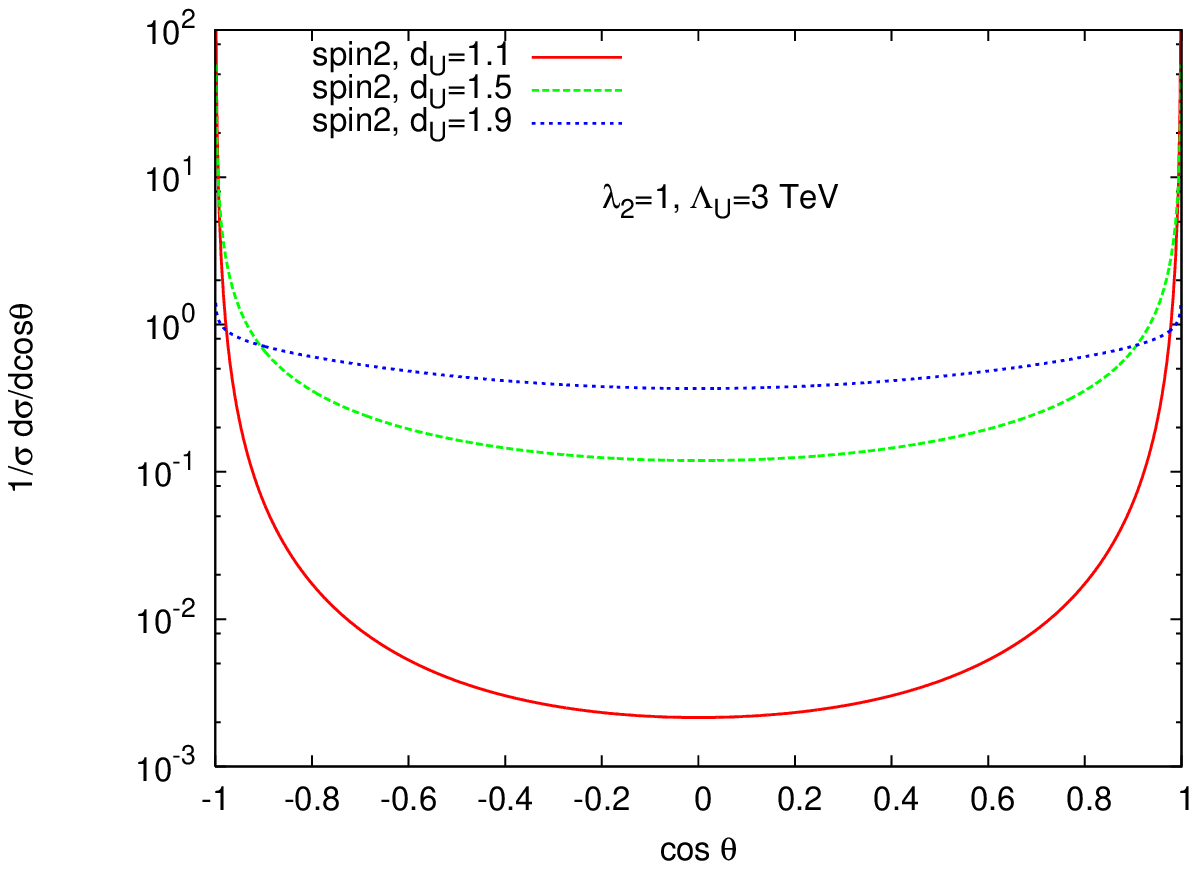}
\caption{\small \label{norm-angular} 
Normalized angular distributions of 
$\gamma\gamma \to \gamma\gamma$ via spin-0 and spin-2 unparticle
exchanges for various $d_\U$ at $\sqrt{s_{\gamma\gamma}} = 0.5$ TeV.
}
\end{figure}

In Fig.\ref{norm-angular}, we show the normalized 
angular distributions $d\sigma /d\cos\theta$
at $\sqrt{s_{\gamma\gamma}} = 0.5$ TeV for spin-0 and spin-2
unparticle exchanges.  Note that in the part for spin-0 the scale on the
$y$-axis is linear while that for spin-2 the scale is logarithmic.
Therefore, in general the spin-2 exchange will give much larger 
contributions in the forward region.  Another interesting feature is that
for spin-0 case when $d_\U$ increases from $1.1$ to $1.9$ the distribution
is becoming more forward.  This is because the factors of $|t|$ and $|u|$
only appear in the numerator, and so when $d_\U$ increases, more powers of
$|t|$ and $|u|$ are contributing in the forward region.  On the other hand,
for spin-2 case more powers of $|t|$ and $|u|$ appear in the denominator
as $d_\U$ is closer to $1$.  Thus, the distribution is much more forward
for small $d_\U$. In fact, it diverges at $|\cos\theta|=1$ for $d_\U <2$.  
We have also verified that the term containing the factor
$\cos(d_\U \pi)$ is affecting the distribution.  
If there were no such a factor, the distribution would have been different,
especially for small $d_\U$.  
This demonstrates the effect of the peculiar 
phase associated with the $s$-channel propagator only.  
If the phase were associated with 
all $s,t,u$ propagators, the effect would have been canceled out when 
we squared the amplitude.

\begin{figure}[t!]
\centering
\includegraphics[width=3.2in]{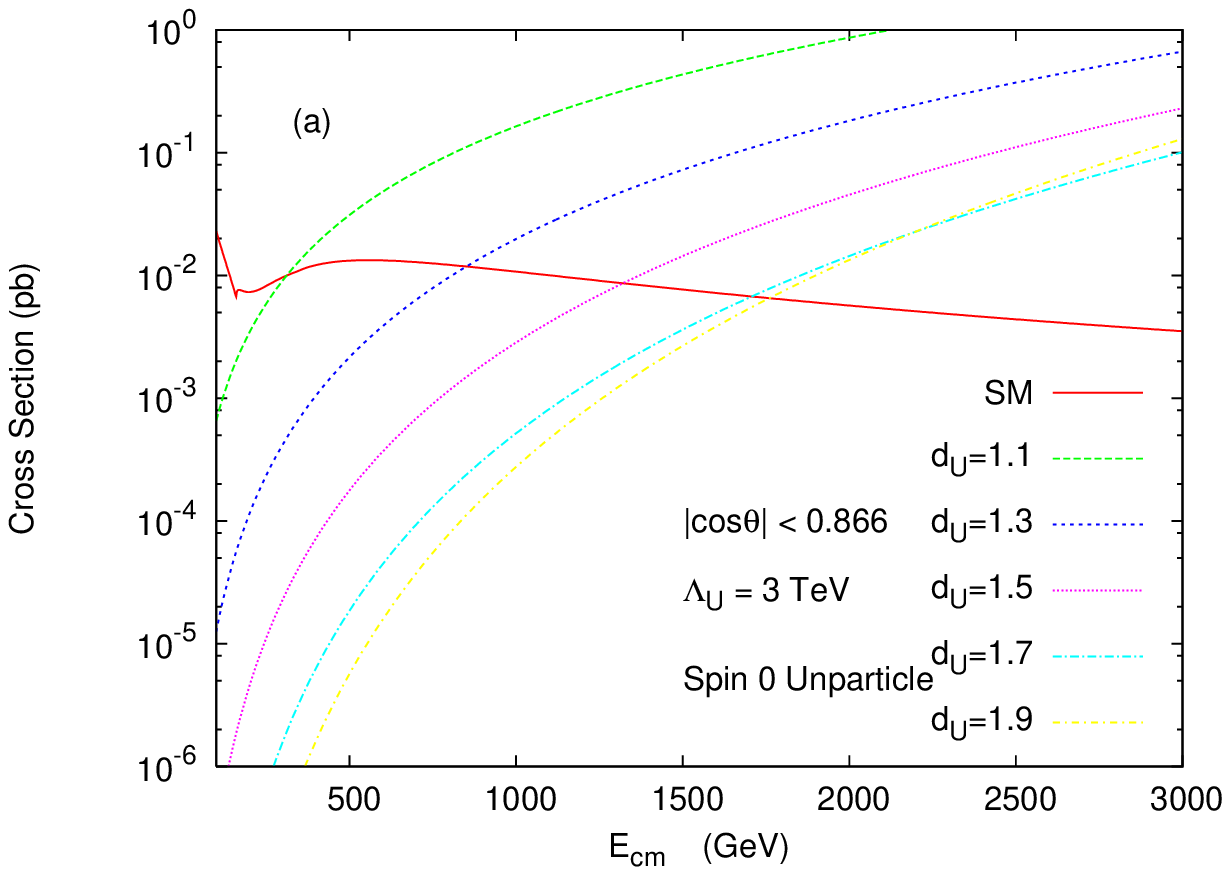}
\includegraphics[width=3.2in]{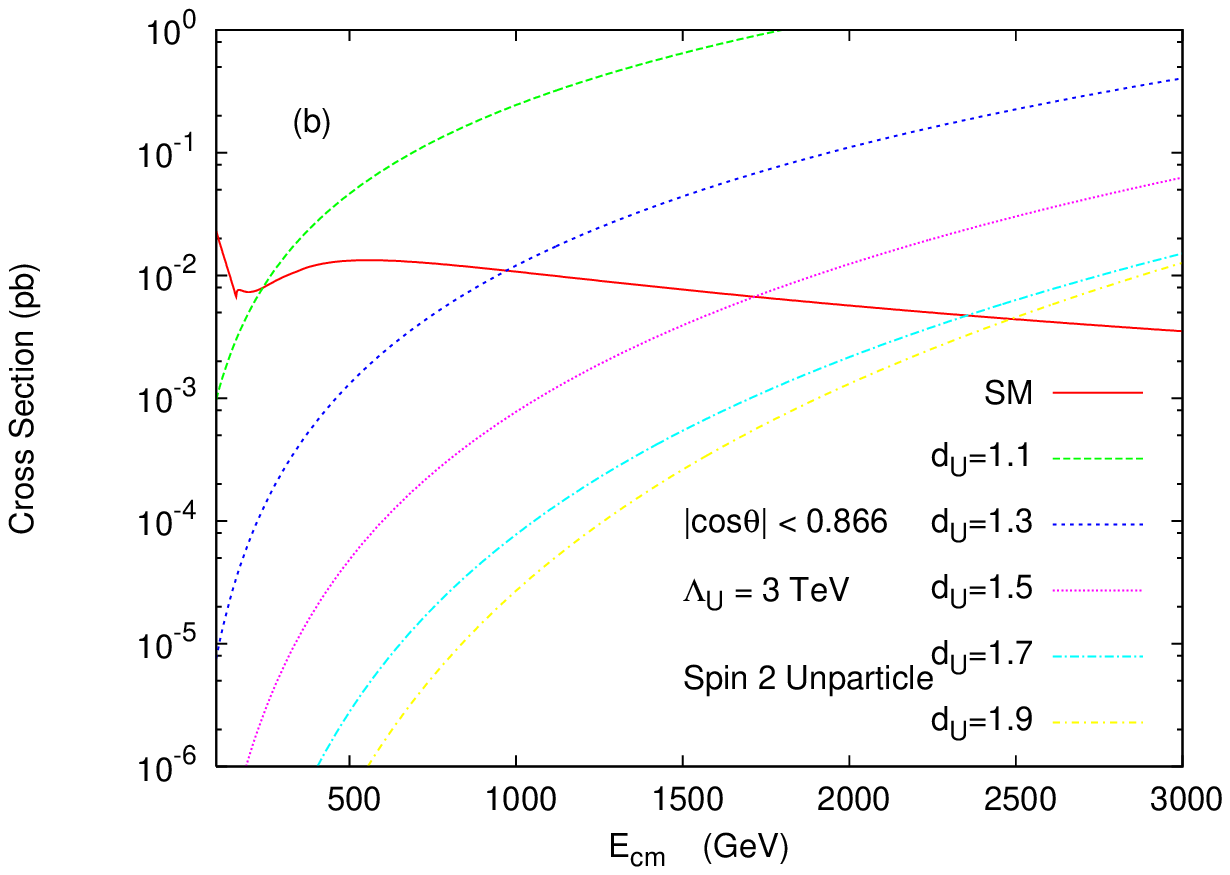}
\caption{\small \label{cross} 
Total cross sections for $\gamma\gamma
\to \gamma\gamma$ via (a) spin-0 and (b) spin-2 unparticle exchange
versus center-of-mass energy for various $d_\U$. The SM expectation is
also shown.  }
\end{figure}

In Fig. \ref{cross}, we plot the integrated cross sections versus the
center-of-mass energy $\sqrt{s_{\gamma\gamma}}$.  We also show the
expectation from the SM, using the results of Ref. \cite{photon} with
the form factors from Ref. \cite{duane}.  
Since the SM cross section peaks in the forward and backward directions, we
impose an angular cut of $|\cos\theta_\gamma| < \cos (30^\circ)$ to reduce 
the SM cross section. 
It is easy to see that the
unparticle cross sections can surpass the SM one at high enough energy
depending on the spin and scaling dimension of the unparticle.
The factor containing $\cos(d_\U \pi)$ also affects the total cross
sections to some extent, especially for small $d_\U$.

\section{Conclusions}

One of the most peculiar features of unparticle is the phase factor
$\exp(-i d_\U \pi)$ associated with the $s$-channel propagator.  We
have studied its effect in $\gamma\gamma \to \gamma\gamma$, which
would have been symmetric in $s$-, $t$-, and $u$-channels without the
phase factor.  However, since the phase is only associated with the
$s$-channel, the effect will show up in the interference terms between
$s$- and $t$- and between $s$- and $u$-channels.  
The effect of such a factor affects the angular distribution and total
cross sections to some significance, especially for small $d_\U$.
We have also shown that the scattering cross sections due to
unparticle exchanges easily surpass the SM contribution.
Thus, the possibility of studying photon scattering in the future linear
collider, using either laser backscattering technique or bremsstrahlung,
is important to test the existence of any tree-level photon-photon scattering.
Unparticle is a unique example that allows tree-level exchange and
contains a special CP-conserving phase factor $\exp(-i d_\U \pi)$ solely in the $s$-channel 
propagator to facilitate interesting interference effects.

\section*{Acknowledgments}
This research was supported in parts by the NSC
under grant number NSC 96-2628-M-007-002-MY3 and the NCTS of Taiwan (Hsinchu).

\end{document}